\documentclass[12pt]{article}
\usepackage {graphicx}
\begin{document}
\title{The Angular-Diameter-Distance-Maximum and Its Redshift as Constraints 
on $\Lambda \neq 0$ FLRW Models }
\author{Marcelo E. de Ara\'{u}jo * and William R. Stoeger **}

\maketitle

\noindent
*Departamento de F\'{i}sica-Matem\'{a}tica, Instituto de F\'{i}sica,
Universidade do Brasil, 21.945-970, Rio de Janeiro, R. J., Brazil \\

\noindent
**Vatican Observatory Research Group, Steward Observatory, 933 North
Cherry Avenue, The University of Arizona, Tucson, AZ 85721

\begin{abstract}
The plethora of recent cosmologically relevant data has indicated that our
universe is very well fit by a standard Friedmann-Lema\^{i}tre-Robertson-Walker
(FLRW) model, with $\Omega_{M} \approx 0.27$ and $\Omega_{\Lambda} \approx
0.73$ -- or, more generally, by nearly flat FLRW  models with parameters close
to these values. Additional independent cosmological information, particularly
the maximum of the angular-diameter (observer-area) distance and the redshift
at which it occurs, would improve and confirm these results, once sufficient
precise Supernovae Ia data in the range $1.5 < z < 1.8$ become available. We
obtain characteristic FLRW closed functional forms for $C = C(z)$ and
$\hat{M}_0 = \hat{M}_0(z)$, the angular-diameter distance and the density per
source counted, respectively, when $\Lambda \neq 0$, analogous to those
we have for $\Lambda = 0$. More importantly, we verify that for flat FLRW
models $z_{max}$ -- as is already known but rarely recognized -- the redshift
of $C_{max}$, the maximum of the angular-diameter-distance, uniquely gives
$\Omega_{\Lambda}$, the amount of vacuum energy in the universe, independently
of $H_0$, the Hubble parameter. For non-flat models determination of both
$z_{max}$ and $C_{max}$ gives both
$\Omega_{\Lambda}$ and $\Omega_M$, the amount of matter in the
universe, as long as we know $H_0$ independently. Finally,
determination of $C_{max}$ automatically gives a very simple observational
criterion for whether or not the universe is flat -- presuming that it is
FLRW. 
\end{abstract}

\section{Introduction}
Over the last 10 or 12 years a great deal of outstanding observational
work has indicated that the best fit model of our universe is a nearly
flat Friedmann-Lema\^{i}tre-Robertson-Walker (FLRW) model with $\Omega_M 
\approx 0.27$ and $\Omega_{\Lambda} \approx 0.73 $ (Riess {\it et al.} 1998;
Perlmutter {\it et al.} 1999; Bennett {\it et al} 2003 (WMAP results);
Peacock {\it et al.} 2001; Percival {\it et al.} 2001; Efstathiou {\it et al.}
2002; Spergel {\it et al.} 2003, and references therein),
where $\Omega_M$ and $\Omega_{\Lambda}$ are the usual density parameters for 
matter, including nonbaryonic dark matter, and dark energy, modelled here
as vacuum energy (the cosmological constant $\Lambda$), respectively. Here
and throughout this paper $\Omega_M$ and $\Omega_{\Lambda}$ refer to these
quantities as evaluated at our time now. This
remarkable concordance is based on WMAP cosmic microwave background
(CMB) anisotropy measurements, a large number of Supernovae Ia
data (see Riess {\it et al.} 2004), and large scale structure studies, and has
been confirmed by other more recent work. Riess and his collaborators (Riess
{\it et al.} 2004), for instance, have recently found a best-fit cosmology 
having $\Omega_M = 0.29$ and $\Omega_{\Lambda} = 0.71$ for their sample of
16 distant ($z > 1$) SN Ia, including 6 with $z > 1.25$, assuming the
universe is exactly flat. Within
the errors this is consonant with the ``concordance'' model given above. \\

Despite the strength of these results, they will obviously have to undergo
gradual
revision and continual verification, as more precise data from higher redshifts
are acquired. When $\Lambda \neq 0$, there are at present, from a strictly
mathematical consideration of the Einstein field equations, not yet enough
completely independent observables to constrain all the free parameters
of the cosmological model (Hellaby, 2006; Stoeger \& Hellaby, in
preparation). \\  

Assuming that the universe is spherically symmetric on the largest scales
(FLRW or, more generally, Lema\^{i}tre-Tolman-Bondi (LTB)), one generally needs
redshifts, luminosity distances (or angular-diameter distances), and galaxy
number counts, together with a reliable galaxy evolution model, or an 
equivalent set of measurements, to constrain the model fully (see Ellis, {\it
et al.} 1985). If $\Lambda \neq 0$, however, or if there is
some other form of dark energy, these data are not enough. We need at least
one other independent parameter -- that is, independent of the observables we
have just mentioned and therefore of those which depend upon them. And,
strictly speaking, this is what we have not had. Thus, the impressive fittings
that have led to the concordance model are still model-dependent in some sense. \\

There is another pair of such independent observables. These would
improve and verify our cosmological fitting, when we are able to obtain an
adequate number of precise luminosity distances -- or angular-diameter 
distances -- and redshifts for SN Ia, or for other standard candles or standard
rods , out to $z \approx 1.8$. These observables are the maximum of the angular-diameter distance (or observer-area distance) $C_{max}$ and the
redshift $z_{max}$ at which it occurs. It has been realized for many years
(McCrea 1935, Hoyle 1961, Ellis \& Tivon 1985) that this distance reaches
a maximum for relatively low redshifts in FLRW universes. For an
Einstein-deSitter ($ \Omega =1 $)universe filled with matter, for instance,
the observer area distance C has a maximum $C_{max}$ at $z_{max} = 1.25$. This
effect is due to the global gravitational focusing of light rays
caused by the matter in the universe -- in effect the entire universe, filled
with homogeneously distributed matter, acts like a gravitational lens. \\

Krauss and Schramm (1993) recognized that, for flat FLRW universes, 
determination of $z_{max}$ would give us $\Omega_{\Lambda}$. They plotted
and provided a table giving this unique correspondence (see their Table 1),
and proposed the possibility of using the measurement of compact parsec-scale
radio jets to observationally exploit it, if the source-evolution problem
can be tamed. Since then, there has been little
development or discussion of this potentially important connection -- except
for Hellaby's (2006) recent closely connected exploration of such
measurements within the more general context of LTB universes (see below). 
Certainly, it is implicit in the Friedmann equation -- most clearly in 
Refsdal, {\it et al.}'s (1967) numerical results of general cosmological
models, in the brief treatment of cosmic distances by Carroll, {\it et al.},
1992 (see pages 510-512, and their Figure 5), and in Peeble's treatment of
angular diameters in cosmology (Peebles 1993), but not pointed out or
discussed further, until Hellaby's more general treatment. This may be
partially due to the
difficulty of obtaining reliable data at the redshifts where we would expect to
locate $C_{max}$ (see below). Now, however, there is the very real prospect of
obtaining angular diameter distances (indirectly, by measuring luminosity
distances of SN Ia) out to $z \approx 1.8$ using telescopes in space. Thus, it
is important to point out again and stress this promising connection, which could
eventually be incorporated in the Bayesian-Fisher matrix (see, for example,
Albrecht, {\it et al.}, 2006) fitting of models to data, or be used as an
independent consistency check on such fittings. \\

Recently, as already mentioned, Hellaby (2006) emphasized the importance
of such a measurement within
a more general framework. He points out that in any LTB cosmology with
$\Lambda = 0$ (which includes all $\Lambda = 0$ FLRW cosmologies as special
cases) the measurement of $C_{max}$ is equivalent to a measurement of the
total mass $M_{max}$ within the sphere defined by $C_{max}$. For
$\Lambda \neq 0$ we have for any LTB model, instead, a simple relationship
between the $\Lambda$, $C_{max}$ and $M_{max}$ (see equation (11) below). So
a measurement of $M_{max}$, or its equivalent, and $C_{max}$ determines
$\Lambda$. What becomes apparent is that $C_{max}$ and the redshift $z_{max}$
at which it occurs constitute independent cosmological observables -- directly
constraining $\Lambda$ and $\Omega_M$ (see Hellaby's Figure A1 in his Appendix,
which shows how different cosmologcal parameters vary with $z_{max}$.)  \\

Applying this directly to flat FLRW models, like those we have good evidence
represent our universe, we quickly see that, since we implicitly have a 
relation between the total mass-energy density and the matter density, or
equivalently between the matter density and $\Omega_{\Lambda}$ --- i.e.
$\Omega_M = 1 - \Omega_{\Lambda}$ --- observational determination of 
$z_{max}$ will directly determine $\Omega_{\Lambda}$ in a very simple and
straightforward way, supporting Krauss and Schramm's results (1993). In
this paper we shall integrate and generalize these results, first of all
verifying Krauss and Schramm's results for flat FLRW universes and writing
down that relationship as an algebraic equation in closed form (they
presented their results numerically), and then generalizing those results
to non-flat FLRW universes, using the relationship Hellaby (2006) noticed.
In this case, $C_{max}$ and $z_{max}$ directly determine both
$\Omega_{\Lambda}$ and $\Omega_M$, if we know $H_0$ independently.
In the course of doing this, we shall, as useful and important by-products,
obtain the FLRW $C = C(z)$ and $\hat{M}_0 = \hat{M}_0(z)$ closed-form functional
relationships for $\Lambda \neq 0$ universes, parallel to those which
are well-known for $\Lambda = 0$ FLRW models (Ellis and Stoeger 1987;
Stoeger, {\it et al.} 1992), as well as a very simple observational
criterion for flatness in terms of $C_{max}$. Here $C(z)$, of course, is
simply the angular-diameter distance as a function of the redshift $z$, and 
$\hat{M}_0(z)$ is the mass density per source counted as a function of
$z$, which is closely related to the differential galaxy number counts
$dN/dz$ (see Stoeger, {et al.} 1992). To our knowledge, these more
general results, along with the closed-form expressions and the flatness
criterion are new. \\

We have already indicated that these measurements will be able to be
implemented once we have luminosity distances and redshifts for SN Ia, or
for other standard candles or standard rods, in the interval $1.5 < z < 1.8$.
As we shall show, it is precisely in this region that a flat FLRW universe will
have a maximum in its angular-diameter distance, if $0.59 \leq \Omega_{\Lambda}
< 0.82$. For the best fit FLRW given by Riess {\it et al.} (2004) with
$\Omega_M = 0.29$ and $\Omega_{\Lambda} = 0.71$, $z_{max} = 1.62$. Another 
potential way of obtaining such precise measurements is -- following Krauss
and Schramm's (1993) idea -- the use of VLBI to determine the
angular-size/redshift relation for ultra-compact (milliarcsecond) radio
sources. These have been argued to be standard rods (Jackson and Dodgson
1997; Jackson 2004).  If we actually do 
find the maximum angular-diameter distance near this value of the redshift,
this would be independent confirmation of the concordance model. If we
do not, but find the maximum angular-diameter distance $C$ at some other value
of $z$, this will require further work at reconciling the models, and
possibly modifying them. In that case, either the universe may still be flat,
but the relative amounts of matter and dark energy would be quite different
from that given by the concordance, or there is a significant deviation from
flatness that must be taken into account, or possibly there are significant
deviations from FLRW on the largest scales which must be included -- or all
three! At the very least, this would be a good consistency check on  
our cosmological fitting so far. Alternatively, as we have already mentioned,
we could simply include both $C_{max}$ and $z_{max}$ data in our over-all 
fitting scheme -- which would further improve the relibility of our results. \\

It is important to point out that this redshift range is already attracting 
special attention. That is because there have been preliminary indications
(Gilliland {\it et al.} 1999) from an SN Ia at $z \approx 1.7$ that the
universe was decelerating at that time! Further studies (Riess {\it et al.}
2001; Mortsell {\it et al.} 2001; Ben\'{i}tez {\it et al.} 2002) have confirmed
the plausibility of that conclusion, but were unable to strengthen it without
further SN Ia measurements in that interval. Thus, we now have two strong
motivations for pursuing precise SN Ia searches and measurements in this
redshift range. \\
 
Finally, one might wonder how measurements of the luminosity distances
of SN Ia can reveal maxima in the angular-diameter (or observer-area)
distances. The luminosity distances themselves will not have such maxima. The
answer to this question is simple, though rarely adverted to. According the
reciprocity theorem of Etherington (1933; see also Ellis 1971), the luminosity
distance $d_L$ is very generally related to the angular-diameter, or
observer-area, distance by

\begin{equation}
d_L = (1+z)^2 C.  \label{recth}
\end{equation}

\noindent
This simple but important relationship holds for all cosmologies, even very
inhomogenous ones. Thus, with observed luminosity distances and redshifts in the
above mentioned crucial redshift range, we can very quickly convert to
angular-diameter distances, and determine whether the maximum for those 
distances lies within that range. \\

Now we shall go on to work out the simple relationship between $z_{max}$
and $\Omega_{\Lambda}$ for flat FLRW. 

\section{The Maximum Angular-Diameter Distance in Flat FLRW with $\Lambda \neq
0$}

The basic equations relating $z_{max}$ and $\Omega_{\Lambda}$ in  flat FLRW
with $\Lambda \neq 0$ are not difficult, but require some effort to obtain
and check, because they involve elliptic integrals. As we have already 
mentioned, this represents the simplest and clearest example of a more general
relationship between the redshift of the maximum of the angular-diameter
distance (in LTB models this is often referred to as the ``areal radius'') and
the matter and vacuum-energy content of the universe for all FLRW and LTB
models (Hellaby 2006). Furthermore, neither Krauss and Schramm (1993) nor
Hellaby (2006) illustrate the actual calculation. Their results were obtained
numerically, and presented in plotted or table form. \\

In flat FLRW, the angular-diameter (or observer-area) distance $C(\eta, y)$ is
given by 

\begin{equation}
C(\eta, y) = R(\eta)y = \frac{R_0 y}{1 + z}\>, \label{oadef}
\end{equation}

\noindent
where $R(\eta)$ is the scale factor, $\eta$ is the conformal time, $R_0$ is
the scale factor now, $y$ is the comoving radial coordinate, and $z$ is the
redshift of signals from distant sources. Here we have used the important
FLRW relationship

\begin{equation}
1 + z = \frac{R_0}{R(\eta)}\>. \label{red}
\end{equation}

Clearly, if we differentiate equation (\ref{oadef}) with respect to $y$ and set
the result equal to zero, we
shall have the equation for the maximum of $C(\eta, y)$, subject to the 
usual condition that $d^2C/dy^2 < 0$ for $dC/dy = 0$. We have then from
equation  (\ref{oadef})

\begin{equation}
dC/dy = \frac{R_0}{1+z} - \frac{R_0 y}{(1+z)^2}dz/dy = 0\,, \label{dcdy}
\end{equation}
which becomes

\begin{equation}
\frac{R_0}{1+z} - \frac{R_0 y}{(1 + z)^2} R_0 H_0 \sqrt{\Omega_{\Lambda} 
+ (1 - \Omega_{\Lambda})(1+z)^3} = 0\,, \label{dcdy2}
\end{equation}
\noindent
since the Friedmann equation in this case yields

\begin{equation}
dz/dy = R_0 H_0 \sqrt{\Omega_{\Lambda} + (1 - \Omega_{\Lambda})(1+z)^3}\>.
                                 \label{dzdy}
\end{equation}

Thus, from solving equation (\ref{dcdy2}) for $y$, we obtain the equation for
$y_{max}$, the comoving radial coordinate distance to the point down
the observer's past light cone at which the angular-diameter distance is a
maximum, as a function of $z_{max}$, the redshift there, and of
$\Omega_{\Lambda}$:

\begin{equation}
y_{max} = \frac{1 + z_{max}}{R_0H_0 \sqrt{\Omega_{\Lambda} + (1+z_{max})^3
                                    (1-\Omega_{\Lambda})}} \> . \label{ymax}
\end{equation}

This is the first and most essential step in finding the equation we are
looking for. \\

The second step involves finding the explicit solution to the Friedmann
equation, essentially equation (\ref{dzdy}), to give us another expression for
$y_{max}$ at $z_{max}.$ Substituting this expression into left-hand-side of
equation (\ref{ymax}) gives a unique implicit equation for $\Omega_{\Lambda}$ as a
function simply of $z_{max}$. This is the relationship we have been looking
for. \\

So, what is the solution of equation (\ref{dzdy})? Normally, we might want to simply
do a numerical integration. However, this would not be very useful in
our case. It turns out, as is well known (Byrd \& Friedman (1954), pp.
8-10 and formula 260.00 (p. 135); see also Jeffrey (1995), pp. 225-226), that,
since this equation involves the square 
root of a cubic polynomial, it has an analytic solution in terms of elliptic
integrals. In our case the most useful form of the solution is:

\begin{equation}
y = \frac{g}{R_0 H_0 \Omega_{\Lambda}^{1/2}}\biggl[F(\phi, k) \mid_{(1+z)^{-1}=1} -
                      F(\phi, k)\mid_{(1+z)^{-1}}\biggr] \>, \label{yeliptic}
\end{equation}

\noindent
where the $F(\phi, k)$ are standard elliptic integrals of the first kind,
for the angle $\phi$, which is a function of $1+z$, and $k$ is the modulus.
More explicitly 

\begin{eqnarray}
 \phi &=&cos^{-1}\Biggl[\frac{-m(1+z)+ (\sqrt 3 -1)}{-m(1+z)-(\sqrt 3 + 1)}\Biggr]\>, \nonumber  \\
 m &= &\Biggl[\frac{1 - \Omega_{\Lambda}}{\Omega_{\Lambda}}\Biggr]^{1/3}, \nonumber \\
k^2 &=&\frac{1}{2} + \frac{\sqrt 3}{4}\>, \nonumber  \\
g &=&\frac{1}{3^{1/4}}\Biggl[\frac{\Omega_{\Lambda}}{1 - \Omega_{\Lambda}}\Biggr]^{1/3}. \nonumber 
\end{eqnarray}

\noindent
This solution was obtained and checked using elliptic integral tables
in Byrd \& Friedman (1954) (entry 260.00, p. 135) in conjunction with
MAPLE.

With equation  (\ref{yeliptic}) being substituted for $y$, equation (\ref{oadef}) is the characteristic
FLRW relationship for the angular-diameter distance $C = C(z)$ in terms of
$z$. It turns out (see below) that this same form of the relationship
holds in the general (non-flat) FLRW cases -- with the parameters $\phi$,
$k$, and $g$ being more complicated functions, involving $\Omega_{\Lambda}$,
either $\Omega_M$ or $C_{max}$, and $H_0$. We shall explicitly write these
down in the next section. Similarly, we quickly can write down the
complementary characteristic $\Lambda \neq 0$ mass density per source counted
as a function of $z$ (see Ellis and Stoeger 1987 and Stoeger, {\it et al.}
1992):

\begin{equation}
 \hat{M}_0(z) = \frac{\mu_{m_{0}}(1+z)^2}{R_0H_0\sqrt{\Omega_{\Lambda} +\Omega_{M} (1+z)^3 - (\Omega_0 - 1)(1+z)^2}}, \label{Mz}
\end{equation}
where $\mu_{m_{0}}$ is the mass-energy density now and
$\Omega_0\equiv\Omega_{\Lambda} +\Omega_M$, and the last term under the
radical sign in the denominator is zero when the universe is flat (see
below). These characteristic FLRW relationships for $C(z)$ and for
$\hat{M}_0(z)$ are very useful to know (Ellis and Stoeger 1987;
Stoeger, {\it et al.}(1992). If the universe is FLRW and $\Lambda = 0$,
then these relationships inevitably follow. If, on the other hand, the data
can be put into these functional forms, then it can be shown by solving
the field equations with this data (Stoeger, {\it et al.} 1992;
Ara\'{u}jo, Stoeger, {\it et al.}, in preparation) that the universe
must be FLRW. Thus, being able to fit the data to these forms, assures us
that the universe is FLRW. Not being able to do so, assures us that it is
not FLRW.
 
Returning to the main object of our derivation, substituting equation (\ref{yeliptic}) into
the left-hand-side of equation  (\ref{ymax}),
we have simply:

\begin{eqnarray}
\frac{g}{\Omega_{\Lambda}^{1/2}}\biggl[F(\phi, k)\mid_{(1+z)^{-1} = 1} &-& F(\phi, k)
  \mid_{(1+z_{max})^{-1}}\biggr]    \nonumber \\
& & = \frac{1+z_{max}}{\sqrt{\Omega_{\Lambda} + (1+z_{max})^3(1 - \Omega_{\Lambda})}} \>.  \label{elzmax}
\end{eqnarray}

This is a transcendental relationship for $\Omega_{\Lambda}$ as a
function of $z_{max}$. It is worth noticing that it does not involve
any other parameters! This is the relationship which represents the
numerical results obtained by Krauss and Schramm (1993). \\

The solutions to this implicit algebraic equation were obtained using
MAPLE, and were checked by hand for values of $\Omega_{\Lambda}$ near
the concordance model value of $\Omega_{\Lambda} = 0.73$. They are given in 
Table 1 and Figure 1 below.\footnote{There are alternative sequences of
steps for obtaining these results -- for instance using the solution of (6) to
write down a general formula for $C$ as a function of $z+1$ and then
differentiating this, setting the result to zero, and solving for
$\Omega_{\Lambda}$ in terms of $z_{max}$. But they all involve explicitly or
implicitly the steps we have indicated -- solving the Friedmann equation to
obtain the relationship between $y$ and the observable $z$ (redshift), and
determining the equation for $C_{max}$ in terms of $y_{max}$ or, from the first
step, its observational equivalent $z_{max}$. Because of the complication of
including a non-zero $\Omega_{\Lambda}$, at some point a numerical solution
will always be needed. See, for instance Carroll, {\it et al.} (1992), pp.
510-512. We have chosen to keep the solution of Friedman
equation analytic, in terms of elliptic integrals, in order to derive the
characteristic FLRW closed-form expression for $C(z)$ and to solve the resulting
algebraic equation numerically.} We can immediately see, that for the
concordance model we should find $z_{max} = 1.64$. For the nearby best fit
model of Riess,
{et al.} (2004) we have already mentioned, $z_{max} = 1.62$. Values of
$z_{max}$ for many other values of $\Omega_{\Lambda}$ are given, as well. These
verify the values presented in Krauss and Schramm (1992), and those
evident in the plots of Refsdal, {\it et al.} (1967), Carroll, {\it et al.}
(1992), and Hellaby (2006). \\

\section{Non-Flat FLRW Universes}

If the universe is not flat, a slight
generalization of these same equations obtains, with the solution for $y$
taking the same general form as given in equation  (\ref{yeliptic}). The generalization
of equation (\ref{elzmax}) in this case will, however, include -- as is intuitively
clear -- a dependence on $\Omega_M$ as well as on $\Omega_{\Lambda}$. Using the
general relationship emphasized by Hellaby (2006)

\begin{equation}
 \Lambda C_{max}^3 - 3C_{max} + 6 M_{max} = 0,  \label{lcmmax}
\end{equation}

\noindent
we can determine $\Omega_M$ through $M_{max}$ in terms of $C_{max}$ and 
$\Lambda$. It is important to stress that equation (\ref{lcmmax}) holds for these 
quantities as measured at $z_{max}$, or $y_{max}$, down the observer's past
light cone. From Hellaby's (2006) results, we easily find that, for FLRW,

\begin{equation}
M_{max} = \frac{4}{3} \pi \rho_M C_{max}^3, \label{masseq}
\end{equation}
where $\rho_M = \rho(t_{max}) = \rho_0(1+z_{max})^3.$ Here $\rho_0$ is
the density at our time now, $t_0$.
Using this together with the definition of $\Omega_M \equiv 8\pi\rho_0/3{H_0}^2$
and equation (\ref{lcmmax}), we easily obtain\footnote{As in Hellaby (2006), we also use units such that $G=c=1$.}

\begin{equation}
\Omega_M = \frac{1}{H_0^2(1+z_{max})^3}[C_{max}^{-2} - \Omega_{\Lambda}H_0^2]. \label{omegam}
\end{equation}

This can be substituted into the non-flat versions of equations (\ref{dzdy}) and (\ref{ymax}),

\begin{equation}
dz/dy = R_0H_0 \sqrt{\Omega_{\Lambda} + \Omega_M(1+z)^3 - (\Omega_0 -1)
   (1+z)^2}, \label{dzdy2}
\end{equation}
and
\begin{equation}
y_{max} = \frac{1+z_{max}}{R_0H_0\sqrt{\Omega_{\Lambda}+\Omega_M(1+z_{max})^3
  - (\Omega_0 - 1)(1+z_{max})^2}}, \label{ymax2}
\end{equation}

In passing, we immediately see from equation (\ref{omegam}) that we have a 
useful observational criterion for flatness of an FLRW universe:

\begin{equation}
\Omega_0 = 1 \Rightarrow (1+z_{max})^{-3}\Biggl[\frac{1}{H_0^2C_{max}^2} - \Omega_{\Lambda}\Biggr] + 
  \Omega_{\Lambda} - 1 = 0, \label{flatness}
\end{equation}
Thus, if already know that the universe is flat, or nearly so, and we know both $z_{max}$ and $C_{max}$, we
can directly determine $\Omega_{\Lambda}$, and therefore $\Omega_M$ itself
from equation (\ref{flatness}).

Proceeding on, then, equation (\ref{omegam}) can therefore be substituted into the
non-flat version of equation (\ref{elzmax}), which
is the same as equation (\ref{elzmax}), except that its right-hand-side is identical
to right-hand-side of equation (\ref{ymax2}) without the $R_0H_0$ factors
in the denominator (these have cancelled out, as before). Thus, we have, finally,
the resulting algebraic relationship involving  $C_{max}$, $z_{max}$,
$H_0$ and $\Omega_{\Lambda}$ as the general FLRW relationship corresponding to the flat case
given in equation (\ref{elzmax}):

\begin{eqnarray}
&  &\frac{g}{\Omega_{\Lambda}^{1/2}}\biggl[ F(\phi, k) \mid_{(1+z)^{-1} = 1} -F(\phi, k) \mid_{(1+z_{\max})^{-1}}\biggr]  \nonumber \\
&  & \hfill{\qquad}= \frac{1+z_{\max}}{\sqrt{\Omega_{\Lambda} + \Omega_{M}(1+z_{\max})^3 -(\Omega_0-1)(1+z_{\max})^2}}.  \label{omegaleq}
\end{eqnarray}
Here and in the solution of the Friedmann  equation for the general
FLRW case, the parameters associated with that solution are now given by:

\begin{eqnarray}
\phi_{(1+z)^{-1}} &=&  \cos^{-1}\Biggl[\frac{(A-B) - (\bar{A}+\bar{B})A(1+z)}
{(A+B)-(\bar{A}+\bar{B})A(1+z)}\Biggr], \nonumber \\
k^2 &=& \frac{(A+B)^2 - (a - b)^2}{4AB}, \nonumber \\
g &=& \frac{1}{\sqrt{AB}}, \nonumber
\end{eqnarray}
with $a \equiv -\frac{\Omega_0-1}{\Omega_{\Lambda}}$,
$b \equiv \frac{\Omega_M}{\Omega_{\Lambda}}$,
and

\begin{eqnarray}
A^2 &=& \bar{A}^2 + \bar{B}^2 - \bar{A}\bar{B}, \nonumber \\
B^2 &=& 3(\bar{A}^2 + \bar{B}^2) + 3\bar{A}\bar{B}. \nonumber 
\end{eqnarray}

Here, further, 

\begin{eqnarray}
\bar{A} = \Biggl\{\frac{\Omega_M}{2\Omega_{\Lambda}} +
\Biggl[\frac{{\Omega_M}^2}{4\Omega_{\Lambda}^2} -
\frac{(\Omega_0-1)^3}{27 \Omega_{\Lambda}^3}\Biggr]^{1/2}\Biggr\}^{1/3},  \nonumber \\
\bar{B} = \Biggl\{\frac{\Omega_M}{2\Omega_{\Lambda}} -
\Biggl[\frac{{\Omega_M}^2}{4\Omega_{\Lambda}^2} -
\frac{(\Omega_0-1)^3}{27 \Omega_{\Lambda}^3}\Biggr]^{1/2}\Biggr\}^{1/3}. \nonumber 
\end{eqnarray}
In these equations, remember that $\Omega_M$ is given by equation (\ref{omegam}), so that relationship given by
equation (\ref{omegaleq}) is indeed an algebraic relationship involving $C_{max}$,
$z_{max}$, $H_0$ and $ \Omega_{\Lambda}$.
Thus, if both $C_{max}$ and $z_{max}$, together with $H_0$, are all known from data, then
equation (\ref{omegaleq}) will determine $\Omega_{\Lambda}$, the only unknown. Using 
that result in equation (\ref{omegam}) will also determine $\Omega_M$. Thus,
observational determination of both $C_{max}$ and $z_{max}$,
will determine both $\Omega_M$ and $\Omega_{\Lambda}$, as long as we also
know  $H_0$. \\

\clearpage
\begin{table}
\begin{tabular} {c c c c c c c c } \hline
${\bf \Omega_{\Lambda}}$&${\bf z_{max}}$&${\bf \Omega_{\Lambda}}$&${\bf z_{max}}$&${\bf \Omega_{\Lambda}}$&${\bf z_{max}}$&${\bf \Omega_{\Lambda}}$&${\bf z_{max}}$ \\ 
0.59&1.50&0.65&1.55&{\bf0.71}&{\bf1.62}&0.77&1.71 \\ 
0.60&1.51&0.66&1.56&0.72&1.63&0.78&1.72 \\ 
0.61&1.51&0.67&1.57&{\bf0.73}&{\bf1.64}&0.79&1.74 \\ 
0.62&1.52&0.68&1.58&0.74&1.66&0.80&1.76 \\ 
0.63&1.53&0.69&1.59&0.75&1.67&0.81&1.78 \\ 
0.64&1.54&0.70&1.61&0.76&1.69&0.82&1.81 \\ \hline
\end{tabular}
\caption{List of pairs ($\Omega_{\Lambda}$,$z_{max}$) for $0.59 \leq \Omega_{\Lambda}
\leq 0.82$ and $1.5 \leq z_{max} \leq 1.81$.}
\end{table}
\clearpage
\begin{figure}
\begin{center}
\includegraphics {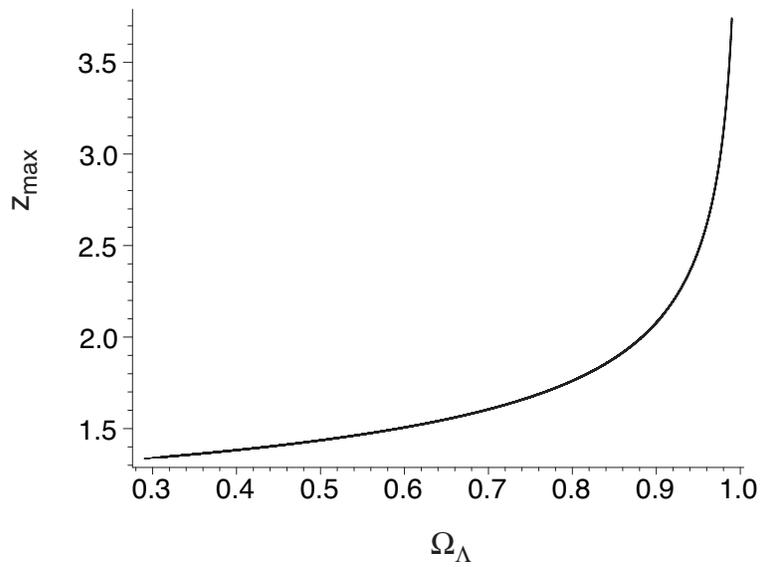}
\end{center}
\caption{Plot of $\Omega_{\Lambda}$ -- $z_{max}$, given by equation (\ref{elzmax}), which
is for a flat FLRW universe. Here $z_{max}$ is the
redshift at which the maximum of the angular diameter distance, $C_{max}$
occurs.}
\label{araujo}
\end{figure}
\clearpage

\section{Observational Prospects and Conclusion}

What are the prospects for actually determining $C_{max}$ and $z_{max}$ from
observations? We would
certainly need precise SN Ia luminosity-distance, or ultra-compact radio-source
angular-diameter distance, and redshift data out to
$z \approx 1.8$ or so. In the SN Ia case this would require careful,
long-range programs
using space-telescopes. However, as already mentioned, we already have
detected and measured SN Ia out to $z \approx 1.7$, and in a recent
assessment (Davis, Schmidt and Kim 2006), precision SN Ia measurements
to $z \approx 1.8$ are considered attainable. This is already considered
an important goal, in order to confirm at what redshift (and cosmic epoch)
the universe made the transition from deceleration to acceleration. It is
certainly fortuitous that the same redshift range promises to provide a
strong independent test of the concordance FLRW model we have derived from
CMB, SN Ia, and large-scale structure measurements. \\

Here we have provided a brief presentation of the straightforward relationship
(first found in numerical form by Krauss and Schramm (1992)) between the
present value of $\Omega_{\Lambda}$ and the redshift $z_{max}$ at
which the angular-diameter (or observer area) distance $C$ occurs in a flat
FLRW cosmology, like that which apparently models our universe. Furthermore,
we have generalized this to non-flat FLRW cases, adding the $C_{\max}$
measurements themselves. In doing this we have derived the characteristic
FLRW observational relationships in closed form for $C(z)$ and $\hat{M}_0
(z)$ in the $\Lambda \neq 0$ case, and found a very simple and potentially
useful observational criterion for flatness.  These results
promise to provide improved determination of the
best fit cosmological model, or a strong consistency test of it, (depending on
how the relationship and the data supporting it are used), once we have
enough precise high-redshift luminosity-distance (or angular-diameter distance)
data. That should be possible in the near future with the rapid progress being
made in SN Ia measurements from space. If the concordance model --
a nearly flat universe with $\Omega_M = 0.27$ and $\Omega_{\Lambda} = 0.73$ --
is approximately correct, we should find observationally that $z_{max}
\approx 1.64$. \\ 

Our thanks to George Ellis and Charles Hellaby for discussions and comments,
and to an anonymous referees for several helpful suggestions for clarification 
and for checking our results, and to one of them for pointing out the
much earlier 1993 Krauss and Schramm paper.  \\

\noindent
{\Large \bf References } \\

\noindent
Albrecht, A., {\it et al.}, 2006, Report of the Dark Energy Task Force, 
astro-ph/0609591. \\
Ben\'{i}tez, N., Riess, A. G., Nugent, P., Dickinson, M., Chornook, R., \& 
Filippenko, A. V. 2002, ApJ, 577, L1. \\
Bennett, C. L., {\it et al.} 2003, ApJS, 148, 1. \\
Byrd, P. F. \& Friedman, M. D. 1954, {\it Handbook of Elliptic
Integrals for Engineers and Physicists}, Springer Verlag. \\
Carroll, S. M.,  Press, W. H., \& Turner, E. L., 1992, ``The Cosmological
Constant,'' Ann. Rev. Astron. \& Astrophys. 30, 499-542.  \\
Efstathiou, G., {\it et al.} 2002, MNRAS, 330, L29 \\
Davis, T. M., Schmidt, B. P. \& Kim, A. G. 2006, PASP, 118, 205. \\
Ellis, G. F. R. 1971, ``Relativistic Cosmology,'' in {\it General Relativity
and Cosmology}, Proc. Int. School Phys. ``Enrico Fermi,'' R. K. Sachs,
editor (New York: Academic Press), pp. 104-182 (see especially pp. 153-1540.\\
Ellis, G. F. R., Nel, S. D., Maartens, R., Stoeger, W. R., \& Whitman, A. P.
1985, Phys. Reports, 124 (No. 5 and 6), 315. \\
Ellis, G. F. R. \& Tivon, G. 1985, Observatory, 105, 189.\\
Ellis, G. F. R. \& Stoeger, W. R. 1987. Class. Quantum Grav., 4, 1697. \\
Etherington, I. M. H. 1933, Phil. Mag., 15, 761. \\
Gilliland, R. L., Nugent, P. E., \& Phillips, M. M. 1999, ApJ, 521, 30. \\
Hellaby, C. W. 2006, MNRAS, 370, 239 (astro-ph/0603637). \\
Hoyle, F., 1961, in Moller, C., ed., Proc. Enrico Fermi School of Physics,
Course XX, Varenna, {\it Evidence for Gravitational Theories}, Academic
Press, New York, p. 141. \\
Jackson, J. C. \& Doddgson, M. 1997, Mon. Not. R. Astron. Soc., 285, 806. \\
Jackson, J. C. 2004, JCAP, 11, 007. \\
Jeffrey, A., 1995, {\it Handbook of Mathematical Formulas and Integrals},
Academic Press, Inc., pp. 225-234. \\
Krauss, L. M., and Schramm, D. N. 1993, ApJ, 405, L43. \\
McCrea, W. H., 1935, Z. Astrophys., 9, 290. \\
Mortsell, E., Gunnarson, C., \& Goobar, A. 2001, ApJ 561, 106. \\
Peacock, J. A., {\it et al.} 2001, Nature, 410, 169. \\
Peebles, P. J. E., 1993, {\it Principles of Physical Cosmology}, Princeton
University Press, Princeton, NJ, pp. 325-329. \\
Percival, W. J., {\it et al.} 2001, MNRAS, 327, 1297. \\
Perlmutter, S., {\it et al.} 1999, ApJ, 517, 565. \\
Refsdal, S., Stabell, R., \& de Lange, F. G. 1967, Mem. R. Astron. Soc., 71,
143. \\
Riess, A. G., {\it et al.} 1998, AJ, 116, 1009. \\
Riess, A. G., {\it et al.} 2001, ApJ, 560, 49. \\
Riess, A. G., {\it et al.} 2004, ApJ, 607, 665. \\
Spergel, D. N., {\it et al.} 2003, ApJS, 148, 175.\\
Stoeger, W. R., Ellis, G. F. R. \& Nel, S. D. 1992, Class. Quantum Grav., 9,
509.  
\end{document}